\documentclass[twocolumn,aps,floatfix]{revtex4}
\usepackage{latexsym,amsmath,amssymb,amsbsy}
\usepackage{graphicx}
\usepackage{enumerate}
\usepackage{nccfoots}
\usepackage[symbol]{footmisc}

\begin{document}

\title{Coherent Dissociation of Relativistic $^{12}$N Nuclei}

\author{\textbf{R.~R.~Kattabekov$^{\textbf{1), 3)}}$, K.~Z.~Mamatkulov$^{\textbf{1), 2)}}$, S.~S.~Alikulov$^{\textbf{2)}}$, D.~A.~Artemenkov$^{\textbf{1)}}$, R.~N.~Bekmirzaev$^{\textbf{2)}}$, V.~Bradnova$^{\textbf{1)}}$, P.~I.~Zarubin$^{\textbf{1)~*}}$, I.~G.~Zarubina$^{\textbf{1)}}$, N.~V.~Kondratieva$^{\textbf{1)}}$, N.~V.~Kornegrutsa$^{\textbf{1)}}$, D.~O.~Krivenkov$^{\textbf{1)}}$, A.~I.~Malakhov$^{\textbf{1)}}$, K.~Olimov$^{\textbf{1), 3)}}$, N.~G.~Peresadko$^{\textbf{4)}}$, N.~G.~Polukhina$^{\textbf{4)}}$, P.~A.~Rukoyatkin$^{\textbf{1)}}$, V.~V.~Rusakova$^{\textbf{1)}}$, R.~Stanoeva$^{\textbf{1), 5)}}$, and S.~P.~Kharlamov$^{\textbf{4)}}$}
\Footnotetext{1)}{Joint Insitute for Nuclear Research, Dubna, Moscow Region, 141980 Russia}
\Footnotetext{2)}{A. Kodirii Jizzakh State Pedagogical Institute, st. S.Rashidov 4, Jizzakh, 130114 Republic of Uzbekistan}
\Footnotetext{3)}{Institute for Physics and Technology, Uzbek Academy of Sciences, ul. G. Mavlyanova 2b, Tashkent, 700084 Republic of Uzbekistan}
\Footnotetext{4)}{Lebedev Physical Institute, Russian Academy of Sciences, Leninskii pr. 53, Moscow, 119991 Russia}
\Footnotetext{5)}{South–West University, Ivan Michailov str. 66, 2700 Blagoevgrad, Bulgaria}
\Footnotetext{*}{E-mail: \texttt{zarubin@lhe.jinr.ru}}}

\indent \par
\noindent \affiliation{Received March 21, 2012}

\begin{abstract}
\noindent  \textbf{Abstract}—The dissociation of relativistic $^{12}$N nuclei having a momentum of 2 GeV$/$c per nucleon and undergoing the most peripheral interactions in a track emulsion is studied. The picture of charged topology of product ensembles of relativistic fragments and special features of their angular distributions are presented.\par

\indent \par
\noindent \textbf{DOI:} 10.1134$/$S1063778813100074
\end{abstract}

\maketitle

\indent The track-emulsion method has still retained its exceptional position as a means for studying the structure of fragmentation of relativistic nuclei owing to the completeness of observation of fragment ensembles and owing to its record spatial resolution. The objective of employing a track emulsion in the Becquerel project [1] at the nuclotron of the Joint Institute for Nuclear Research (JINR, Dubna) is to study the clustering of nucleons for a wide variety of light nuclei, including radioactive ones. An analysis proves to be the most comprehensive for coherent-dissociation events, in which target fragments or mesons are not produced [2] and which, for the sake of brevity, are referred to as white stars. Since the perturbation is minimal in interactions of this type, a configuration overlap of the ground states of the nuclei being studied and the observed ensembles of fragments and nucleons manifests itself. On this basis, there arise possibilities for some kind of a tomography of the nuclear structure. The distribution of white stars with respect to the probability for the formation of various configurations shows a correlation with weights of respective cluster components of the nuclei being studied. The discovery of the contribution from unexplored and even unexpected components of deeply bound cluster states is possible on this path. This is so especially for proton-rich light nuclei. The picture of nucleon clustering in events of the coherent dissociation of the $^{12}$N radioactive nucleus, which has yet to receive adequate study, on track-emulsion nuclei is the subject of the present investigation. This investigation is the next step in studying the cluster structure of the $^{7}$Be [3], $^{8}$B [4], and $^{9}$C [5] radioactive nuclei. The role of the $^{12}$N nucleus in nuclear astrophysics is that it continues the sequence of these nuclei in proton-pickup reactions in nucleosynthesis and ensures an alternative scenario of the synthesis of the $^{12}$C isotope. Thus, the structure of these nuclei may manifest genetic connections, and this is why an investigation of these nuclei within an unified approach is so appealing.\par
\indent For $^{12}$N white stars, one can expect that the channels $^{11}$C + \emph{p} (the threshold is 0.6 MeV), $^8$B $+$ $^4$He (the threshold is 8 MeV), and \emph{p} $+$ $^7$Be $+$ $^4$He, as well as the channels associated with the cluster dissociation of the core nucleus $^{7}$Be, make a leading contribution to the distributions of the fragment charge \emph{Z}$_{fr}$. A feature that distinguishes the coherent dissociation of the $^{12}$N nucleus from the case of lighter nuclei in the vicinity of the proton drip line is that it may receive a contribution from the decays of unbound nuclei of $^8$Be$_{g.s.}$ and $^9$B$_{g.s.}$. In particular, the threshold for the $^3$He $+$ $^9$B$_{g.s.}$ channel is 10 MeV. A small difference in binding energy in relation to channels involving \emph{Z}$_{fr}$ $>$ 2 fragments suggests two types of an interpretation for the $^{12}$N nucleus. One one hand, the bound nuclei of $^7$Be and $^8$B may be assumed to be its core. On the other hand, the role of the core may be played by the unbound $^8$Be$_{g.s.}$ and $^9$B$_{g.s.}$ systems.\par
\indent A track nuclear emulsion was irradiated with a mixed beam of relativistic $^{12}$N, $^{10}$C, and $^{7}$Be nuclei at a beam momentum of \emph{p}$_{0}$ $=$ 2 GeV$/$c per nucleon. The beam was formed upon the charge-exchange process involving primary $^{12}$C nuclei and their fragmentation [6]. The irradiated stack contained 15 layers of track nuclear emulsion BR-2, which possessed sensitivity up to relativistic particles. Each layer had transverse dimensions of 10 $\times$ 20 cm$^2$ and a thickness of about 0.5 mm. The implementation of irradiation was such that the beam propagated parallel to the stack plane along its long side and filled the stack entrance window as uniformly as was possible. An analysis of the charge topology in the coherent dissociation of these nuclei confirmed the dominance of the isotopes $^{7}$Be and $^{10}$C in the beam and the presence of $^{12}$N nuclei there [7]. The choice of charge exchange for the generation of $^{12}$N nuclei was justified by the fact that it was convenient to identify tracks of beam nuclei in the track emulsion. At the same time, this choice bounded sharply statistics of interactions of $^{12}$N nuclei.\par
\begin{figure}
\includegraphics[width=0.45\textwidth]{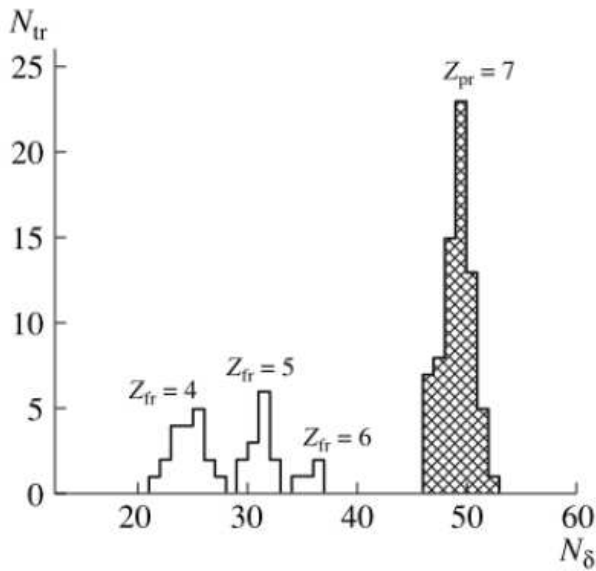}
\caption{\label{Fig:1} Distribution of the number of beam particles (shaded histogram) and secondary fragments (unshaded histograms), \emph{N}$_{tr}$, with respect to the average number of delta electrons,  $<$\emph{N}$_{\delta} >$, in white stars satisfying the conditions \emph{Z}$_{pr}$ $=$ 7 and $\Sigma$\emph{Z}$_{fr}$ $=$ 7.}
\end{figure}
\indent The present analysis is based on scanning without selections 12 layers of the irradiated track emulsion along the tracks of primary particles with charges \emph{Z}$_{pr}$ visually estimated as \emph{Z}$_{pr}$ $>$ 2 over the length of about 1088 m. This resulted in finding 7241 inelastic interactions, including 608 white stars containing only relativistic fragments in the angular cone specified by the inequality \emph{$\theta$}$_{fr}$ $<$ 11$^{\circ}$. In white stars that could be associated with $^{12}$N nuclei, the average values of the delta-electron density \emph{N}$_{\delta}$ were measured along the tracks of beam nuclei and secondary fragments with \emph{Z}$_{fr}$ $>$ 2. The relativistic-particle charges of \emph{Z}$_{fr}$ $=$ 1, 2 are determined visually with a high reliability.\par 
\begin{table}[!ht]
\caption{\label{Tabel:1}Charge-topology distribution of fragments from white stars, \emph{N}$_{ws}$ , where the total charge of relativistic fragments is $\Sigma$\emph{Z}$_{fr}$ $=$ 6, and from $\Sigma$\emph{Z}$_{fr}$ $=$ 6 events, \emph{N}$_{tf}$ , accompanied by target fragments or product mesons}
\begin{center}
\begin{tabular}{l|c|c} \hline
~\ Channel ~\ & ~\ $<$\emph{N}$_{\delta} >$ ~\ & ~\ RMS ~\ \\ \hline
~\ Be (19) ~\ & ~\ 24.0 $\pm$ 0.4 ~\  & ~\ 1.5 $\pm$ 0.3 ~\ \\ 
~\ B (13) ~\ & ~\ 30.6 $\pm$ 0.3 ~\ & ~\ 0.9 $\pm$ 0.2 ~\ \\
~\ C (4) ~\ & ~\ 35.3 $\pm$ 0.4 ~\ & ~\ 0.8 $\pm$ 0.3 ~\ \\
~\ N (72) ~\ & ~\ 48.6 $\pm$ 0.2 ~\ & ~\ 1.4 $\pm$ 0.1 ~\ \\
\hline
\end{tabular}
\end{center}
\end{table}
\indent As was shown in studying the $^8$B [4] and $^9$C [5] nuclei, the application of the condition requiring that relativistic fragments conserve the projectile charge, \emph{Z}$_{pr}$ $=$ $\Sigma$\emph{Z}$_{fr}$, makes it possible to remove the contribution from the charge-exchange process involving lighter accompanying nuclei. The dominance of carbon nuclei in this irradiation rendered the aforementioned selection especially important and justified the application of the cumbersome procedure of counting delta electrons. For these events, Fig.~\ref{Fig:1} shows the distribution of the number of tracks, \emph{N}$_{tr}$, of beam particles that have the identified charge of \emph{Z}$_{pr}$ $=$ 7 and secondary fragments that have the identified charges of \emph{Z}$_{fr}$ with respect to the measured delta-electron density \emph{N}$_{\delta}$ over 1 mm of length. The mean ($<$\emph{N}$_{\delta} >$) and root-mean-square (RMS) values of the delta-electron density that are presented in Table 1 indicate that the charge classification of the tracks was quite reliable. The measurements of the quantities \emph{N}$_{\delta}$ made it possible to select 72 white stars satisfying the conditions \emph{Z}$_{pr}$ $=$ 7 and $\Sigma$\emph{Z}$_{fr}$ $=$ 7.\par
\indent An identification of H and He fragments by their total momentum \emph{p$\beta$c} in a track nuclear emulsion by the method of measuring multiple scattering would be of great interest. However, only within quite a limited volume can such measurements be performed because of the angular spread of respective tracks and because of their large number. They were performed for a random sample of fragments from 2He $+$ 3H and 3He $+$ H configurations (see Fig.~2). For H fragments, the distribution of \emph{p$\beta$c}$_{H}$ has a mean value of $<$\emph{p$\beta$c}$_{H} >$  $=$ 1.9 $\pm$ 0.1 GeV and an RMS value of 0.2 GeV; this corresponds to the values expected for protons. For He fragments, there are two components of \emph{p$\beta$c}$_{He}$; the first has a mean value of $<$\emph{p$\beta$c}$_{^{3}He} >$ $=$ 5.2 $\pm$ 0.1 GeV and RMS $=$ 0.4 GeV, while the second has a mean value of $<$\emph{p$\beta$c}$_{^{4}He} >$ $=$ 7.2 $\pm$ 0.1 GeV and RMS $=$ 0.3 GeV. They are expected for the relativistic isotopes $^3$He and $^4$He, respectively. The distribution is indicative of approximately equal proportions of the isotopes $^3$He and $^4$He, as might have been expected for the $^{12}$N nucleus.\par
\indent The identification of tracks by charge permits reconstructing the charge topology of white stars generated by $^{12}$N nuclei (see Table 2). On the basis of these data, the contribution of $^{12}$N nuclei to the beam was estimated at a level of 14\% (without allowance for H and He nuclei). According to the accumulated sample of white stars generated by $^{10}$C and $^7$Be nuclei, the contribution of each of these isotopes is about 43\%. For \emph{Z}$_{fr}$ $>$ 2 isotopes, the mass number \emph{A}$_{fr}$ is also determined from \emph{Z}$_{fr}$. For a further selection of coherent-dissociation events featuring only fragments of $^{12}$N nuclei (not involved in the interaction process), the condition for the angular cone was toughened to become $\theta_{fr}$ $<$ 6$^{\circ}$. The value on the right-hand side of this inequality was determined by a lenient constraint on the momentum of the Fermi motion of nucleons. In the distribution of 45 selected events (Table 2), the fraction of channels involving \emph{Z}$_{fr}$ $>$ 2 heavy fragments reaches about 2$/$3, but the contribution of channels featuring only light fragments (He and H) remains quite significant. The sample of events associated with the 2He $+$ 3H channel proved to be unexpectedly large. Taking into account the fact that, in the dissociation of the $^7$Be nucleus, the branching fractions of the 2He and He $+$ 2H channels are approximately equal to each other [3] and assuming the $^7$Be core in the $^8$B [4] and the $^9$C [5] nuclei, one would expect an approximate equality of the branching fractions of the 2He $+$ 3H and 3He $+$ H channels for the $^{12}$N nucleus.\par
\begin{figure}
\includegraphics[width=0.4\textwidth]{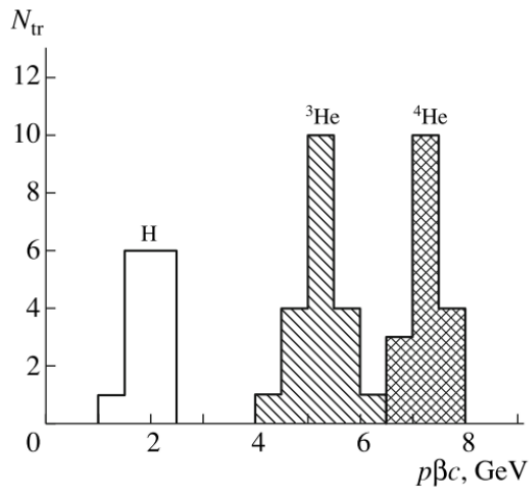}
\caption{\label{Fig:2} Distribution of measured values of \emph{p$\beta$c} for (unshaded histogram) H and (shaded histograms) He fragments from the 2He $+$ 3H and 3He $+$ H configurations.}
\end{figure}
\begin{table}
\caption{\label{Tabel:2}Distribution of the number of white stars (\emph{N}$_{ws}$) among the dissociation channels where the total charge of fragments is $\Sigma$\emph{Z}$_{fr}$ $=$ 7 and where the measured charge of the beam track is \emph{Z}$_{pr}$ $=$ 7}
\begin{center}
\begin{tabular}{c|c|c} \hline
\multicolumn{1}{c|}{Channel} 
& \multicolumn{2}{|c}{\emph{N}$_{ws}$} \\ \cline{2-3}
& \multicolumn{1}{|c|}{ \emph{$\theta$}$_{fr} <$ 11$^{\circ}$ } & \multicolumn{1}{|c}{ \emph{$\theta$}$_{fr}$ $<$ 6$^{\circ}$ } \\ \hline
~\ He $+$ 5H ~\ & ~\ 9 ~\  & ~\ 2 ~\ \\ 
~\ 2He $+$ 3H ~\ & ~\ 24 ~\ & ~\ 12 ~\ \\
~\ 3He $+$ H ~\ & ~\ 2 ~\ & ~\ 2 ~\ \\
~\ $^7$Be $+$ 3H ~\ & ~\ 10 ~\ & ~\ 5 ~\ \\
~\ $^7$Be $+$ He $+$ H ~\ & ~\ 9 ~\ & ~\ 8 ~\ \\
~\ $^8$B $+$ 2H ~\ & ~\ 11 ~\ & ~\ 9 ~\ \\
~\ $^8$B $+$ He ~\ & ~\ 3 ~\ & ~\ 3 ~\ \\
~\ C $+$ H ~\ & ~\ 4 ~\ & ~\ 4 ~\ \\ \hline
~\ $\Sigma$ ~\ & ~\ 72 ~\ & ~\ 45 ~\ \\ \hline
\end{tabular}
\end{center}
\end{table}
\indent Let us now consider the results obtained by measuring fragment-production angles \emph{$\theta$} in 45 white stars bounded by the fragmentation cone \emph{$\theta$}$_{fr} <$ 6$^{\circ}$. By way of example, Fig.~3 shows the measured polar angles \emph{$\theta$} of all fragments in events featuring \emph{Z}$_{fr}$ $>$ 2 fragments. The mean and root-mean-square (RMS) values of the distributions of \emph{Z}$_{fr}$ $>$ 2 fragments with respect to the polar angle \emph{$\theta$} are given in Table 3. Measurements of the angles \emph{$\theta$} make it possible to estimate the fragment transverse momenta \emph{p$_{T}$} and their sums according to the approximation \emph{P$_{T}$} $\approx$ \emph{A$_{fr}$P$_{0}$}~sin~\emph{$\theta$}.\par
\indent In particular, one can estimate the mean transverse momentum in C $+$ H events, which are interpreted as $^{11}$C $+$ \emph{p}. The threshold for the similar channel $^{10}$C $+$ \emph{d} is substantially higher, amounting to 11.5 MeV. The mean value of $<$\emph{P$_{T}$}~($^{11}$C $+$ \emph{p})$>$ $=$ 300 $\pm$ 52 MeV$/$c, together with RMS $=$ 104 $\pm$ 37 MeV$/$c, is indicative of the nuclear-diffraction type of the dissociation process in question [8] and of the presence of the bounce-off effect. The branching fraction of the $^{11}$C $+$ \emph{p} channel proved to be (9 $\pm$ 5)\%. This value is close the branching fraction of the reaction $^9$C $\rightarrow$ $^8$B $+$ \emph{p} [5], but it differs significantly from its counterpart in the case of the $^8$B nucleus [4], for which the channel involving the separation of a loosely bound proton, $^7$Be $+$ \emph{p}, has a branching fraction of (50 $\pm$ 12)\%.\par
\begin{figure}
\includegraphics[width=0.4\textwidth]{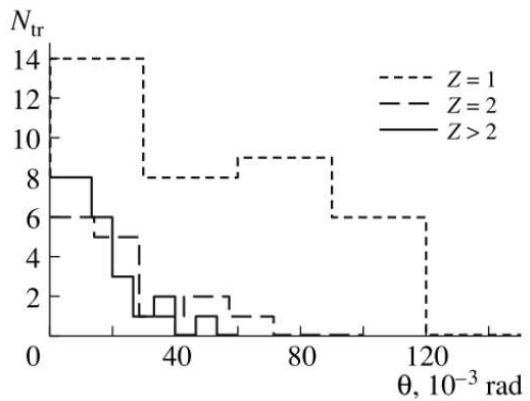}
\caption{\label{Fig:3} Distribution of fragments originating from the coherent dissociation of $^{12}$N nuclei and having charge numbers \emph{Z} with respect to the polar emission angler $\theta$ in events involving \emph{Z}$_{fr}$ $>$ 2 fragments.}
\end{figure}
\indent The contribution of decays from the ground state of the unstable nucleus $^8$Be$_{g.s.}$ to the production of \emph{Z}$_{fr}$ $=$ 2 fragments (see Table 2) presents a serious problem. Owing to an extremely small decay energy for $^8$Be$_{g.s.}$, it becomes possible to sidestep the problem of identifying relativistic isotopes of He. At the momentum of 2 GeV$/$c per nucleon, the decays $^8$Be$_{g.s.}$ $\rightarrow$ 2$\alpha$ of relativistic nuclei are identified by requiring that final-state alpha-particle pairs be within the region of extremely small divergence angles $\Theta_{2\alpha}$ that is specified by the condition $\Theta_{2\alpha} <$ 10.5 $\times$ 10$^{-3}$ rad [4]. In addition, the decays of the $^8$Be nucleus from the first excited state, whose spin–parity is 2$^+$ , were identified by the angles of divergence in the region of 15 $\times$ 10$^{-3} < \Theta_{2\alpha} <$ 45 $\times$ 10$^{-3}$ rad [9$\--$12]. In studying the dissociation of relativistic $^{10}$C nuclei, the decays of $^8$Be$_{g.s.}$ and $^9$B$_{g.s.}$ nuclei were identified [10$\--$12].\par

\indent In the case of the $^{12}$N nucleus, two candidates for the decay of the $^8$Be nucleus from the ground state of spin parity 0$^+$ , which have a divergence angle smaller than 10.5 $\times$ 10$^{-3}$ rad, were found in the distribution of the divergence angles $|Theta$(He $+$ He) for the 2He $+$ 3H and 3He $+$ H white stars (see Fig.~4). On this basis, the contributions of $^8$Be nuclei was estimated at a level of (4 $\pm$ 2)\%. For the neighboring nuclei of $^{12}$C [2], $^{10}$C [10$\--$12], $^{10}$B [13], and $^{14}$N [14], it was about 20\%.\par
\begin{table}
\caption{\label{Tabel:3}Mean and root-mean-square (RMS) values of the distributions of \emph{Z}$_{fr}$ $>$ 2 fragments with respect to the polar angle $\theta$ (in units of 10$^{-3}$ rad)}
\begin{center}
\begin{tabular}{c|c|c} \hline
~\ Channel ~\ & ~\ $<$\emph{$\theta$}$_{Z>3}>$ ~\ & ~\ RMS ~\ \\ \hline
~\ $^7$Be $+$ 3H ~\ & ~\ 33 $\pm$ 14 ~\  & ~\ 31 $\pm$ 10 ~\ \\ 
~\ $^7$Be $+$ He $+$ H ~\ & ~\ 19 $\pm$ 4 ~\ & ~\ 13 $\pm$ 3 ~\ \\
~\ $^8$B $+$ 2H ~\ & ~\ 20 $\pm$ 3 ~\ & ~\ 10 $\pm$ 2 ~\ \\
~\ $^8$B $+$ He ~\ & ~\ 7 $\pm$ 1 ~\ & ~\ 3 $\pm$ 1 ~\ \\
~\ $^{11}$C $+$ p ~\ & ~\ 17 $\pm$ 3 ~\ & ~\ 6 $\pm$ 2 ~\ \\
\hline
\end{tabular}
\end{center}
\end{table}
\indent The data sample in the region of $\Theta$(He $+$ He) smaller than 10.5 $\times$ 10$^{-3}$ rad does not rule out the possibility of dissociation through the 2He channel via the decay of the $^8$Be nucleus from the first excited state, whose spin–parity is 2$^+$. However, the method that we employed for the 0$^+$ ground state and which consisted in selecting pairs of tracks diverging within a narrow cone is no longer applicable. In the $\Theta$(He $+$ He) range being considered, an admixture of $^3$He nuclei should be present, especially in the case of the neutron-deficient nucleus $^{12}$N. Unfortunately, the required level of identification of relativistic helium isotopes remains unattainable at the present time.\par
\begin{figure}
\includegraphics[width=0.4\textwidth]{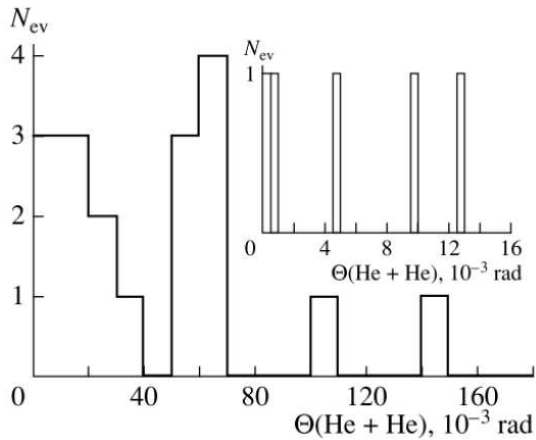}
\caption{\label{Fig:4} Distribution of pairs of He fragments with respect to the angles of divergence, $\Theta$(He $+$ He), for the 2He $+$ 3H and 3He $+$ H white stars. The inset shows an enlarged distribution of $\Theta$(He $+$ He) in the region of the smallest values.}
\end{figure}
\indent In the irradiation run under discussion, we have studied the charge topology of the dissociation of a $^{12}$N nucleus, and this permits characterizing its special features. First, there are no pronounced leading channels, as is suggested by quite a uniform distribution of the data sample among possible channels. In contrast to what we have for the neighboring nuclei of $^{12}$C [2], $^{10}$C [7], and $^{14}$N [14], the copious production of \emph{Z}$_{fr} >$ 3 fragments proceeds in the dissociation of $^{12}$N nuclei. Therefore, the $^7$Be nucleus may play the role of the core in $^{12}$N. In searches for an analogy with the $^9$C nucleus via substituting an alpha-particle cluster for one of the outer protons in the 2\emph{p} $+$ $^7$Be system, there arises the following difficulty. The branching fraction of the channels in which the disintegration of an alpha-particle cluster must occur in the $^{12}$N nucleus is very close to the branching fractions of the channels that may be associated with the separation of an alpha-particle cluster as a discrete unit. In all probability, the \lq\lq simple\rq\rq picture of the $^{12}$N nucleus as the \emph{p} $+$ $^7$Be $+$ $^4$He structure is inadequate. Most likely, the cluster structure of the $^{12}$N nucleus is a complicated mixture of states of the $^7$Be nuclear core and possible configurations of protons and extremely light nuclei.\par

\begin{center}
ACKNOWLEDGMENTS
\end{center}
\indent This work was supported by the Russian Foundation for Basic Research (project no. 12-02-00067) and by grants from the plenipotentiaries of Bulgaria and Romania at the Joint Institute for Nuclear Research.\par

\begin{center}
REFERENCES
\end{center}

\indent 1. The BECQUEREL Project, http:$//$becquerel.jinr. ru \par
\indent 2. V. V. Belaga et al., Phys. At. Nucl. \textbf{58}, 1905 (1995); arXiv: 1109.0817 [nucl-ex].\par
\indent 3. N. G. Peresadko et al., Phys. At. Nucl. \textbf{70}, 1226 (2007); nucl-ex/0605014.\par
\indent 4. R. Stanoeva et al., Phys. At. Nucl. \textbf{72}, 690 (2009); arXiv: 0906.4220 [nucl-ex].\par
\indent 5.  D. O. Krivenkov et al., Phys. At. Nucl. \textbf{73}, 2103 (2010); arXiv: 1104.2439 [nucl-ex].\par
\indent 6. P. A. Rukoyatkin et al., Eur. Phys. J. ST \textbf{162}, 267 (2008).\par
\indent 7. R. R. Kattabekov, K. Z. Mamatkulov, D. A. Artemenkov, et al., Phys. At. Nucl. \textbf{73}, 2110 (2010); arXiv: 1104.5320 [nucl-ex].\par
\indent 8. N. G. Peresad'ko, V. N. Fetisov, Yu. A. Aleksandrov, et al., JETP Lett. \textbf{88}, 75 (2008); arXiv: 1110.2881 [nucl-ex].\par
\indent 9. D. A. Artemenkov et al., Phys. At. Nucl. \textbf{70}, 1222 (2007); nucl-ex/0605018.\par
\indent 10. D. A. Artemenkov et al., Few-Body Syst. \textbf{44}, 273 (2008).\par
\indent 11. D. A. Artemenkov et al., Few-Body Syst. \textbf{50}, 259 (2011); arXiv: 1105.2374 [nucl-ex].\par
\indent 12. D. A. Artemenkov et al., Int. J. Mod. Phys. E \textbf{20}, 993 (2011); arXiv: 1106.1748 [nucl-ex].\par
\indent 13. F. G. Lepekhin, Phys. Part. Nucl. \textbf{36}, 233 (2005).\par
\indent 14. T. V. Shchedrina et al., Phys. At. Nucl. \textbf{70}, 1230 (2007); nucl-ex/0605022.\par

\indent \par
\indent \emph{Translated by A. Isaakyan}

\end{document}